\documentclass[preprintnumbers,showpacs,floats,twocolumn,prd,aps]{revtex4}
\usepackage{amssymb,amsmath}
\usepackage[dvips]{epsfig}

\renewcommand{\vec}[1]{\mbox{\boldmath$#1$}}

\begin{document}

\title{Five-Dimensional Black Hole Capture Cross-Sections}
\author{Cisco Gooding}\email{dwg2@sfu.ca}
\author{Andrei V. Frolov}\email{frolov@sfu.ca}
\affiliation{
  Department of Physics,
  Simon Fraser University\\
  8888 University Drive,
  Burnaby, BC Canada
  V5A 1S6
}
\date{March 6, 2008}

\begin{abstract}
  We study scattering and capture of particles by a rotating black hole in the
  five-dimensional spacetime described by the Myers-Perry metric. The
  equations of geodesic motion are integrable, and allow us to calculate
  capture conditions for a free particle sent towards a black hole from
  infinity. We introduce a three-dimensional impact parameter describing
  asymptotic initial conditions in the scattering problem for a given initial
  velocity. The capture surface in impact parameter space is a sphere for a
  non-rotating black hole, and is deformed for a rotating black hole. We
  obtain asymptotic expressions that describe such deformations for small
  rotational parameters, and use numerical calculations to investigate the
  arbitrary rotation case, which allows us to visualize the capture surface as
  extremal rotation is approached.
\end{abstract}

\preprint{SCG-2007-04}
\maketitle

\section{Introduction}

Recent developments in string theory and brane world models have led to an
increased interest in higher-dimensional solutions of Einstein's field
equations. Of particular importance are higher-dimensional spacetimes
containing black holes, reviewed in Refs.~\cite{Emparan:2008eg,Kanti:2004nr}.
In the context of models with large extra dimensions
\cite{ArkaniHamed:1998rs}, the black hole size could be much smaller than the
effective size of the extra dimensions, and extra dimensions could be (and
often are) treated as having infinite extent. This is the case for mini black
holes that could potentially be created by high-energy collision experiments
in the near future \cite{Banks:1999gd,Eardley:2002re}. As one can expect that
the particle collisions would rarely be head-on, most of such black holes
would be produced in a highly rotating state, which brings a new focus on the
original rotating black hole solution of Myers and Perry \cite{Myers:1986un}.

The properties of higher dimensional rotating black holes have been studied
extensively \cite{Emparan:2008eg,Kanti:2004nr}. It has been recently shown
that the equations of motion in a five-dimensional rotating black hole
background are separable for both particles \cite{Frolov:2003en} and waves
\cite{Frolov:2002xf}. Generalizations to higher dimensions followed quickly
\cite{Frolov:2007nt,Page:2006ka}, with current state of the subject reviewed
in Ref.~\cite{Frolov:2008jr}. Although a proof of integrability formally
solves the problem of geodesic motion in higher-dimensional black hole
spacetimes, properties of the actual geodesics remain largely unexplored. In
this paper, we address this shortcoming using a combination of analytic and
numerical methods.

We study geodesics in a five-dimensional spacetime of a rotating Myers-Perry
black hole \cite{Myers:1986un}. We consider scattering and capture of
particles launched from infinity toward a black hole, and calculate the
capture cross-section and its dependence on black hole rotation. In five
dimensions, the initial conditions at infinity can be described in terms of
three-dimensional impact parameter in addition to the initial particle
velocity. The boundary of a capture region in impact parameter space is a
deformed two-sphere. This is a natural generalization of the usual capture
problem in four dimensions, where the impact parameter has two dimensions, and
the capture surface is a deformed one-sphere (circle)
\cite{1976PhRvD..14.3281Y,Calvani:1981ya,Zakharov:1994ts}.

This paper is organized in the following way: In Section~\ref{sec:eom}, we
review the metric of a five-dimensional rotating black hole, and write down
geodesic equations. Using five existing integrals of motion
\cite{Frolov:2003en}, they can be cast in the form of five first-order
ordinary differential equations. By considering turning points of a ``radial''
equation of motion, we can deduce conditions under which geodesics will be
captured by a black hole. In Section~\ref{sec:impact}, we relate the critical
values of the integrals of motion for such capture to initial conditions at
infinity in terms of the impact parameter. In Section~\ref{sec:results}, we
present our results on capture cross-sections of five-dimensional black holes.
In the limit of a non-rotating black hole, the capture surface is a
two-sphere, the radius of which is easy to find. Deformations due to small
rotational parameters are described analytically by asymptotic expansion of
the equations that define the capture surface. As extremal black hole rotation
is approached, we finally turn to numerical calculations for visualization of
the capture behavior. We finish by discussing our results in
Section~\ref{sec:disc}.

\section{Equations of Motion}\label{sec:eom}

A five-dimensional rotating black hole is described by its mass $M$ and two
independent (dimensionless) rotational parameters $a$ and $b$ that are scaled
by the gravitational radius $r_g$, which is defined such that
\begin{equation}\label{eq:M}
  M \equiv \frac{3\pi r_g^2}{8G},
\end{equation}
where $G$ is the five-dimensional gravitational coupling constant. The
original Myers-Perry metric \cite{Myers:1986un} can be generalized and written
in several different forms, for example in an unconstrained parametrization
\cite{Chen:2006xh}. Here we will follow the notation of Ref.~\cite{Frolov:2003en},
which describes the metric of a 5D rotating black hole in Boyer-Lindquist
coordinates, although we will be using a dimensionless inverse-radius-squared
coordinate $w \equiv (r_g/r)^2$, which proves to be more convenient and leads
to simpler expressions. With this coordinate choice, the metric of a 5D
rotating black hole is written as
\begin{eqnarray}\label{eq:metric}
  ds^2 &=& -dt^2 + \frac{\sigma^2}{w} \left(\frac{dw^2}{4{\cal D}w^2}+d\theta^2\right) + \\
       & & \frac{\alpha}{w}\sin^2{\theta}\,d\phi^2 + \frac{\beta }{w}\cos^2{\theta}\,d\psi^2 + \nonumber\\
       & & \frac{w}{\sigma^2}\left(dt+a\sin^2{\theta}d\phi+b\cos^2{\theta}d\psi\right)^2, \nonumber
\end{eqnarray}
where we have introduced the definitions
\begin{equation}\label{eq:ab}
  \alpha = 1 + a^2 w, \hspace{1em}
  \beta = 1 + b^2 w,
\end{equation}
\begin{equation}\label{eq:D}
  {\cal D} = \alpha\beta - w,
\end{equation}
and
\begin{equation}\label{eq:sigma}
  \sigma^2 = 1 + w\left(a^2\cos^2{\theta}+b^2\sin^2{\theta}\right).
\end{equation}
The spatial coordinates $\{w,\theta,\phi,\psi\}$ are a generalization of
spherical coordinates in Euclidean $3$-space. The two ``azimuthal'' angles
$\phi$ and $\psi$ take on values in the interval $\left[0,2\pi\right]$,
whereas the ``polar'' angle $\theta$ takes on values in the interval
$\left[0,\pi/2\right]$.

The black hole horizons are located at $w=w_\pm$, with
\begin{equation}\label{eq:horizon}
  w_\pm^{-1} = \frac{1}{2}\left[1-a^2-b^2\pm\sqrt{\left(1-a^2-b^2\right)^2-4a^2b^2}\right].
\end{equation}
The two horizons merge and the black hole becomes extremal when
\begin{equation}\label{eq:extremal}
  |a \pm b| = 1.
\end{equation}

Equations of geodesic motion, which describe the trajectories of free
particles, can be derived from the metric by varying the associated
Hamilton-Jacobi action \cite{Frolov:2003en}. As the equations of geodesic
motion for an $N$-dimensional rotating black hole are completely integrable
\cite{Page:2006ka}, they can be cast as a set of first-order nonlinear
ordinary differential equations
\begin{subequations}\label{eq:eom}
\begin{eqnarray}
  \dot{\theta}^2 &=& \Theta, \\
  \dot{w}^2 &=& 4\Upsilon w, \\
  \dot{t} &=& \frac{\sigma^2}{w}\,E + \frac{\alpha\beta}{\cal D}\,{\cal E}, \\
  \dot{\phi} &=& \frac{\Phi}{\sin^2{\theta}}-\frac{a w \beta }{\cal D}\,{\cal E}-\left(a^2-b^2\right) \frac{w \Phi}{\alpha}, \\
  \dot{\psi} &=& \frac{\Psi}{\cos^2{\theta}}-\frac{b w \alpha }{\cal D}\,{\cal E}+\left(a^2-b^2\right) \frac{w \Psi}{\beta}.
\end{eqnarray}
\end{subequations}
Here the over-dots denote differentiation along the trajectory, but we have
abandoned affine parametrization and absorbed a common factor in the geodesic
equations into a parameter redefinition. We have also introduced the
simplifying definitions
\begin{equation}\label{eq:E}
  {\cal E} = E + \left(\frac{a\Phi}{\alpha} + \frac{b\Psi}{\beta}\right) w,
\end{equation}
\begin{equation}\label{eq:Q}
  {\cal Q} = (a^2-b^2) \left(\frac{\Phi^2}{\alpha} - \frac{\Psi^2}{\beta}\right),
\end{equation}
\begin{equation}\label{eq:Upsilon}
  \Upsilon = {\cal D} \left[p^2 - K w + {\cal Q} w^2 \right] + \alpha\beta\, w {\cal E}^2,
\end{equation}
\begin{equation}\label{eq:Xi}
  \Xi = p^2 (a^2 \cos^2\theta + b^2\sin^2\theta),
\end{equation}
and
\begin{equation}\label{eq:Theta}
  \Theta = \Xi - \frac{\Phi^2}{\sin^2\theta} - \frac{\Psi^2}{\cos^2\theta} + K.
\end{equation}
Associated with the geodesic motion are the following conserved quantities:
the particle energy $E$, the momenta $\Phi$ and $\Psi$ that are conjugate to
$\phi$ and $\psi$ (respectively), and an additional integral of motion $K$
(similar to Carter's constant of the 4D Kerr metric) which follows from the
existence of a second-rank Killing tensor of the Myers-Perry metric
\cite{Frolov:2003en}. We have also introduced the flat spacetime scalar
momentum $p$ such that $p^2 = E^2-m^2$.

\begin{figure}
  \centerline{\epsfig{file=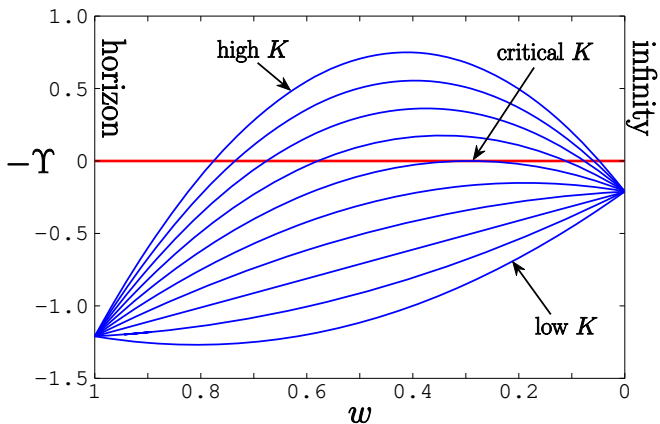, width=246pt}}
  \caption{
    Effective potential $-\Upsilon$ for radial motion of a particle (with $E/m=1.1$)
    around a spherically symmetric 5D black hole, plotted for various
    values of integral of motion $K$ (which plays the role of total
    angular momentum).
    Trajectories with $\Upsilon>0$ (for all $w$) have insufficient
    angular momenta to avoid capture, whereas those with radial
    turning points ($\Upsilon=0$) escape back to infinity. The critical
    $K$ occurs for the curve that contains an unstable circular orbit.
  }
  \label{fig:pot}
\end{figure}

The equations (\ref{eq:eom}) for the radial ($w$) and polar ($\theta$)
directions resemble the energy conservation equation for a particle moving in
an external potential ($-\Upsilon$ and $-\Theta$ correspondingly), and an
understanding of their solutions can be gained from our intuition about such
problems. Of particular interest to us is the radial equation, in which the
existence of turning points determines whether the particle will be captured
or scattered back to infinity. Figure~\ref{fig:pot} shows the effective
potential $-\Upsilon$ (in the case of a non-rotating black hole) for different
values of integral of motion $K$, which plays the role of the total angular
momentum. We see that the value of $K$ determines the height of the potential
barrier the particle has to overcome, and hence whether or not turning points
exist.

\begin{figure}
  \centerline{\epsfig{file=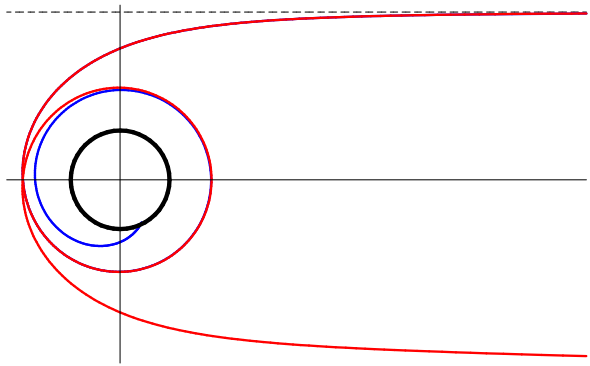, width=220pt}}
  \caption{
    Geodesics in the equatorial plane of a spherically symmetric 5D black hole
    corresponding to Figure~\ref{fig:pot} above. The red one has angular
    momentum slightly above critical, and escapes to infinity, while the blue
    one has angular momentum slightly below critical, and falls into the black
    hole.
  }
  \label{fig:traj}
\end{figure}

The condition $\Upsilon=0$ indicates a radial turning point. The critical
value of the angular momentum $K$ is further distinguished by the condition
that the turning point occurs precisely at the top of the potential barrier,
i.e.\ at $\partial\Upsilon/\partial w=0$. Taken together, these two conditions
describe a location of a circular orbit, which turns out to be unstable in our
case. If we imagine perturbing this orbit, then trajectories with slightly
less angular momentum would result in the particle spiraling into the black
hole, while trajectories with slightly more angular momentum would turn around
and send the particle back to infinity. These two fates of a geodesic are
illustrated in Figure~\ref{fig:traj}, once again for the case of a
non-rotating black hole.

To find the values of the integrals of motion corresponding to the particle
capture threshold, we are then left with the task of solving the system
\begin{equation}\label{eq:crit}
  \Upsilon=0, \hspace{1em}
  \frac{\partial\Upsilon}{\partial w} = 0,
\end{equation}
while eliminating the turning point location $w$. This venture amounts to
finding the $w$-independent condition for a double root of $\Upsilon$ to
exist. Despite giving the impression of a rational function due to appearance
of $\alpha$ and $\beta$ in the denominator of expressions (\ref{eq:E}) and
(\ref{eq:Q}), the effective potential $\Upsilon$ is actually a cubic
polynomial in $w$
\begin{equation}\label{eq:cubic}
  \Upsilon = \gamma_3 w^3 + \gamma_2 w^2 + \gamma_1 w + \gamma_0,
\end{equation}
with the coefficients given by
\begin{subequations}
\begin{eqnarray}
  \gamma_3 &=& (abE+b\Phi+a\Psi)^2 + \\
           & & (a^2-b^2)(b^2\Phi^2-a^2\Psi^2) - K a^2 b^2, \nonumber\\
  \gamma_2 &=& (E+a\Phi+b\Psi)^2-(b\Phi+a\Psi)^2 + \\
           & & (1-a^2-b^2)(K-E^2)+p^2 a^2 b^2, \nonumber\\
  \gamma_1 &=& m^2+p^2(a^2+b^2)-K, \\
  \gamma_0 &=& p^2.
\end{eqnarray}
\end{subequations}
One can calculate the location of the unstable circular orbit by taking a
linear combination of equations (\ref{eq:crit})
\begin{equation}
  3\Upsilon-\frac{\partial\Upsilon}{\partial w}\, w = \gamma_2 w^2 + 2\gamma_1 w + 3\gamma_0 = 0,
\end{equation}
which yields a quadratic equation for orbit radius $w_\star$, with the solution
\begin{equation}
  w_\star = \frac{-\gamma_1 - \sqrt{\gamma_1^2-3\gamma_2\gamma_0}}{\gamma_2}.
\end{equation}
Rather than attempting elimination of $w_\star$ from equation (\ref{eq:crit}),
we will use a better technique. The polynomial (\ref{eq:cubic}) will have a
double root if (and only if) the cubic discriminant
\begin{equation}\label{eq:disc}
  \delta = \gamma_1^2\gamma_2^2-4\gamma_0\gamma_2^3-4\gamma_1^3\gamma_3+18\gamma_0\gamma_1\gamma_2\gamma_3-27\gamma_0^2\gamma_3^2
\end{equation}
vanishes ($\delta=0$). Viewed as an equation for total angular momentum $K$,
this condition defines a critical capture surface in integrals of motion
space, which in principle solves the problem. In practice though, one ends up
with a quartic polynomial in $K$, the solution of which is prohibitively
complex. We will discuss approximate and numerical methods to find the capture
surface in Section~\ref{sec:results}, but before we do that, let us discuss
how integrals of motion relate to initial conditions in the scattering
problem.

\section{Impact Parameter}\label{sec:impact}

Let us now consider a particle launched toward the black hole from very far
away, and figure out how the values of integrals of motion are related to
asymptotic initial conditions. The asymptotic form of the equations of motion
(\ref{eq:eom}) is easy to derive. In the limit $w \rightarrow 0$ we have
\begin{equation}
  \alpha = \beta = {\cal D} = \sigma^2 = 1, \hspace{2em}
  {\cal E} = E, \hspace{2em}
  \Upsilon = p^2,
\end{equation}
and so the asymptotic coordinate velocities are related to the values of
integrals of motion by
\begin{equation}\label{eq:infty:1}
  \frac{\dot{r}}{r_g} = - \frac{p}{w}, \hspace{2em}
  \dot{t} = \frac{E}{w},
\end{equation}
\begin{equation}\label{eq:infty:2}
  \dot{\theta}^2 = \Theta, \hspace{2em}
  \dot{\phi} = \frac{\Phi}{\sin^2 \theta}, \hspace{2em}
  \dot{\psi} = \frac{\Psi}{\cos^2 \theta}.
\end{equation}
As you can see, we chose the trajectory parametrization that leads to constant
coordinate velocities for angular variables at infinity, at the expense of
divergent $\dot{r}$ and $\dot{t}$ (both of which blow up as $1/w$). This does
not affect the proper initial velocity of a particle $v \equiv |\dot{r}|/(r_g
\dot{t}) = p/E$, of course.

If we are far enough away from the black hole, we can approximate the
spacetime as flat (in our case, the 5D Minkowski metric), and we can describe
distances in the four spatial dimensions using the standard Euclidean metric.
Then, the position vector in 4-space can be written in Cartesian coordinates
as
\begin{equation}
  \vec{r} = \left(
    \begin{array}{c}
      x_1 \\
      x_2 \\
      x_3 \\
      x_4 \\
    \end{array}
  \right) =
  \left(
    \begin{array}{c}
      r\sin{\theta}\sin{\phi} \\
      r\cos{\theta}\sin{\psi} \\
      r\sin{\theta}\cos{\phi} \\
      r\cos{\theta}\cos{\psi} \\
    \end{array}
  \right).
\end{equation}
Differentiating the particle position vector with respect to Minkowski time
variable $\tau = r_g t$, we obtain the particle velocity vector $\vec{v}
\equiv \dot{\vec{r}}/(r_g \dot{t})$.

\begin{figure}
  \centerline{\epsfig{file=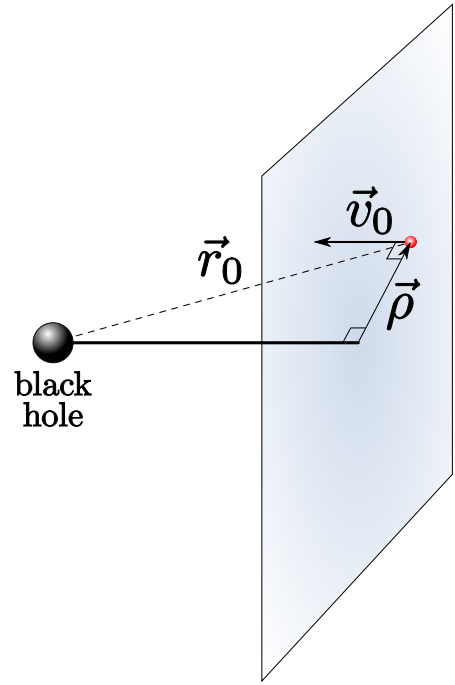, width=2.3in}}
  \caption{
    Geometry of the impact parameter definition in asymptotically flat
    spacetime.
  }
  \label{fig:impact}
\end{figure}

Since the metric described by equation (\ref{eq:metric}) is invariant under
$\phi$ and $\psi$ rotations, we will align our coordinate system such that the
initial values of $\phi$ and $\psi$ are zero. If we denote initial values by
the subscript $0$, then we find that the initial particle position is
\begin{equation}\label{eq:r0}
  \vec{r}_0 = \left(
    \begin{array}{c}
      x_{10} \\
      x_{20} \\
      x_{30} \\
      x_{40} \\
    \end{array}
  \right) = r_0
  \left(
    \begin{array}{c}
      0 \\
      0 \\
      \sin{\theta_0} \\
      \cos{\theta_0} \\
    \end{array}
  \right),
\end{equation}
while the initial particle velocity is directed along
\begin{equation}\label{eq:v0}
  \dot{\vec{r}}_0 = \left(
    \begin{array}{c}
      \dot{x}_{10} \\
      \dot{x}_{20} \\
      \dot{x}_{30} \\
      \dot{x}_{40} \\
    \end{array}
  \right) = \dot{r}_0\, \frac{\vec{r}_0}{r_0} + r_0
  \left(
    \begin{array}{r}
      \dot{\phi}_0\sin{\theta_0} \\
      \dot{\psi}_0\cos{\theta_0} \\
      \dot{\theta}_0\cos{\theta_0} \\
      -\dot{\theta}_0\sin{\theta_0} \\
    \end{array}
  \right).
\end{equation}
Now, along the same lines as the usual four-dimensional analysis
\cite{1976PhRvD..14.3281Y,Calvani:1981ya,Zakharov:1994ts}, we will define the
{\em impact parameter} $\vec{\rho}$ to be the vector in the three-dimensional
hyperplane at infinity orthogonal to initial velocity vector $\vec{v}_0$,
which connects the line of sight to black hole parallel to initial velocity
and the initial particle location. The geometry of this definition is depicted
in Figure~\ref{fig:impact}. From it, we can deduce that our impact parameter
is given by the projection
\begin{equation}
  \vec{\rho} = \vec{r}_0 - \frac{\vec{r}_0\cdot\vec{v}_0}{v_0^2} \, \vec{v}_0.
\end{equation}
Ultimately, we are interested in the limit of a finite impact parameter
infinitely far away from the black hole. In these circumstances, the norm of
the particle velocity is dominated by the first term in equation
(\ref{eq:v0}). Noting that the two terms in equation (\ref{eq:v0}) are
orthogonal, and keeping in mind that the particle is initially descending
inward ($\dot{r}_0\leq0$), one can write the impact parameter as
\begin{equation}
  \vec{\rho} = \frac{r_0^2}{|\dot{r}_0|}
  \left(
    \begin{array}{r}
      \dot{\phi}_0\sin{\theta_0} \\
      \dot{\psi}_0\cos{\theta_0} \\
      \dot{\theta}_0\cos{\theta_0} \\
      -\dot{\theta}_0\sin{\theta_0} \\
    \end{array}
  \right).
\end{equation}
Note that this definition is invariant under trajectory reparametrization, as it should be.

Since the impact parameter vector $\vec{\rho}$ lives in a three-dimensional
hyperplane, we can represent it as $\vec{\rho}=\rho^i \vec{e}_i$ with
$i\in\{1,2,3\}$, where $\{\vec{e}_i\}$ is a set of orthonormal basis vectors
perpendicular to $\vec{v}_0$. A natural choice for this basis is given by
\begin{equation}\label{eq:basis}
  \vec{e}_1 = \left(
    \begin{array}{c}
      1 \\
      0 \\
      0 \\
      0 \\
    \end{array}
  \right),~~
  \vec{e}_2 = \left(
    \begin{array}{c}
      0 \\
      1 \\
      0 \\
      0 \\
    \end{array}
  \right),~~
  \vec{e}_3 = \left(
    \begin{array}{c}
      0 \\
      0 \\
      \phantom{-}\cos{\theta_0} \\
      -\sin{\theta_0} \\
    \end{array}
  \right).
\end{equation}
Although strictly speaking these vectors are orthogonal to $\vec{r}_0$, they
are suitable for our purposes since asymptotically $\vec{r}_0$ and $\vec{v}_0$
become parallel. In the $\{\vec{e}_i\}$ basis, the impact parameter is
represented as
\begin{equation}
  \vec{\rho}
  = \frac{r_0^2}{|\dot{r}_0|}
  \left(
    \begin{array}{c}
      \dot{\phi}_0\sin{\theta_0} \\
      \dot{\psi}_0\cos{\theta_0} \\
      \dot{\theta}_0 \\
      \end{array}
  \right).
\end{equation}
Using the equations of motion evaluated at asymptotic infinity
(\ref{eq:infty:1},\ref{eq:infty:2}), we can relate initial coordinate
velocities to our integrals of motion, in terms of which the impact
parameter vector is written as
\begin{equation}
  \vec{\rho}
  = \frac{r_g}{p}
  \left(
    \begin{array}{c}
      \Phi/\sin{\theta_0} \\
      \Psi/\cos{\theta_0} \\
      \pm\sqrt{\Theta_0} \\
      \end{array}
  \right).
\end{equation}
Taking the Euclidean norm of the above expression, and using definition
(\ref{eq:Theta}) of $\Theta$, we find the length of the impact parameter to be
\begin{equation}\label{eq:impact}
  \frac{\rho^2}{r_g^2} = \left(a^2\cos^2{\theta_0}+b^2\sin^2{\theta_0}\right)+\frac{K}{p^2}.
\end{equation}
Substituting the value of the critical angular momentum $K$ obtained from
equation (\ref{eq:disc}) into the above expression, and observing that
integrals of motion $\Phi$ and $\Psi$ are essentially components $\rho_1$ and
$\rho_2$ of the impact parameter vector, we obtain an equation describing a
two-dimensional capture surface in impact parameter space. The total capture
cross-section is the volume interior to the capture surface.

\section{Capture Cross-Sections}\label{sec:results}

\subsection{Spherical Black Hole}

Let us begin by studying the simplest case first: a non-rotating black hole
($a=b=0$). As the whole spacetime is spherically symmetric, the capture
surface is also a perfect sphere, the radius of which is easy to find. In this
case, the effective potential $\Upsilon(w)$ given by equation
(\ref{eq:Upsilon}) is a simple quadratic
\begin{eqnarray}
  \Upsilon &=& (1-w)(p^2-Kw) + w E^2 \\
           &=& K w^2 + (m^2-K) w + p^2, \nonumber
\end{eqnarray}
as shown in Figure~\ref{fig:pot}. Now, the condition for a double root of
$\Upsilon(w)$ to exist is the vanishing of the discriminant
\begin{equation}
  \left(K-m^2\right)^2-4p^2K=0,
\end{equation}
which implies that the critical value of integral of motion $K$ for capture is
$K_\circ = \left(E+p\right)^2$. The corresponding radius of a sphere in impact
parameter space is
\begin{equation}\label{eq:sph:impact}
  \rho_\circ = \frac{E+p}{p}\, r_g,
\end{equation}
such that all trajectories with smaller impact parameters are captured by the
black hole, and all the ones with the larger impact parameters escape to
infinity. The total capture cross-section is the volume of this
three-dimensional ball of radius $\rho_\circ$
\begin{equation}\label{eq:sph:total}
  \upsilon_\circ = \frac{4\pi}{3} \frac{(E+p)^3}{p^3}\, r_g^3.
\end{equation}
When the black hole is rotating, spherical symmetry is broken, and the capture
surface is deformed away from a sphere. While the exact solution of the
algebraic equations (\ref{eq:crit}), which are in general cubic polynomials,
is too complicated to be of practical use, the deformation can be approximated
analytically quite well for a slowly rotating black hole. Let us discuss two
useful approximations next, and turn to the general case later.

\subsection{First-Order Approximation}

Effects of slow rotation of the black hole on the motion and capture of
particles can be studied by perturbative expansion in the magnitude of the
rotational parameters $a$ and $b$. Expanding the effective potential
$\Upsilon$ to linear order leads to
\begin{equation}
  \Upsilon \simeq \left[K+2E(a\Phi+b\Psi)\right] w^2 + (m^2-K) w + p^2,
\end{equation}
and so once again we are left with a quadratic effective potential, except
that the leading coefficient is perturbed. As in the spherical case, we can
find the critical value of $K$ by setting the discriminant to zero, from which
we obtain
\begin{equation}
  K \simeq K_\circ + 2p(a\Phi+b\Psi).
\end{equation}
The length of the critical impact parameter vector gains a
directionally-dependent term, such that
\begin{equation}
  \rho^2 \simeq \rho_\circ^2 + \frac{2}{p}\left(a\Phi+b\Psi\right) r_g^2.
\end{equation}
This is readily recognizable as a simple origin shift
\begin{equation}
  (\vec{\rho}-\vec{o})^2 \simeq \rho_\circ^2.
\end{equation}
Thus, to linear order in rotational parameters, the capture surface is a
sphere of radius $\rho_\circ$ centered around
\begin{equation}\label{eq:shift}
  \vec{o} = \left(
    \begin{array}{c}
      a \sin\theta_0 \\
      b \cos\theta_0 \\
      0 \\
    \end{array}
  \right) r_g,
\end{equation}
and the total cross-section is unchanged to first order. As we will see later,
this conclusion is confirmed by the general numerical analysis of particle
capture.

\subsection{Ultra-Relativistic Limit}

Second-order accurate analysis of the capture cross-section turns out to be
much harder than the linear one, as order reduction in the effective potential
$\Upsilon$ does not happen. Even after second-order expansion, one is still
left with a quartic equation (\ref{eq:disc}) to solve. However, there is a
special case which allows simplified treatment: the ultra-relativistic limit
($E/m \rightarrow \infty$), for which the order of the discriminant $\delta$
is reduced by one power of $K$. A cubic equation is much easier to solve than
quartic, so after a straightforward but longish calculation (while keeping
only second order terms), we finally obtain
\begin{equation}
  K = (4-a^2-b^2)E^2+2(a\Phi+b\Psi)E - \frac{1}{2} (b\Phi+a\Psi)^2.
\end{equation}
Substituting this into equation (\ref{eq:impact}), and keeping in mind that
for an ultra-relativistic particle $p^2 \simeq E^2$ and $\rho_\circ = 2r_g$,
we obtain a second-order accurate expression for the capture surface given by
\begin{equation}
  (\vec{\rho}-\vec{o})^2 = \rho_\circ^2 - (\vec{\rho}\cdot\vec{s})^2,
\end{equation}
where the origin $\vec{o}$ and the shape distortion $\vec{s}$ are vectors
\begin{equation}
  \vec{o} = \left(
    \begin{array}{c}
      a \sin\theta_0 \\
      b \cos\theta_0 \\
      0 \\
    \end{array}
  \right) r_g,
  \hspace{1em}
  \vec{s} = \frac{1}{\sqrt{2}}\, \left(
    \begin{array}{c}
      b \sin\theta_0 \\
      a \cos\theta_0 \\
      0 \\
    \end{array}
  \right).
\end{equation}
The above equation for the capture surface is a quadratic section, and after
some rearrangement of terms this equation can be brought into the canonical
form
\begin{equation}
  (\vec{\rho}-\vec{o})^T \mathbb{M} (\vec{\rho}-\vec{o}) \simeq \rho_\circ^2,
\end{equation}
where the matrix $\mathbb{M}$ is given by
\begin{equation}
  \mathbb{M} = \mathbb{I} + \vec{s}\otimes\vec{s},
\end{equation}
and we have neglected higher-order terms. It is clear now that the capture
surface is an ellipsoid centered at $\vec{o}$, with interior volume
\begin{equation}
  \upsilon = (\det\mathbb{M})^{-\frac{1}{2}}\, \upsilon_\circ.
\end{equation}
To second order we have $\det\mathbb{M} \simeq 1 + s^2$, and so the total
capture cross-section of ultra-relativistic particles by a (slowly) rotating
black hole is
\begin{equation}\label{eq:approx}
  \upsilon \simeq \left(1 - \frac{a^2}{4}\cos^2\theta_0 - \frac{b^2}{4}\sin^2\theta_0\right) \upsilon_\circ.
\end{equation}
The capture cross-section is maximal for a non-rotating black hole (for which
it is $\upsilon_\circ = 32\pi\,r_g^3/3$) and diminishes when the black hole is
spun up. As we will see, this approximation works remarkably well, even if the
black hole rotation is fast.

\subsection{Numerical Results for General Case}

\begin{figure}
  \centerline{\epsfig{file=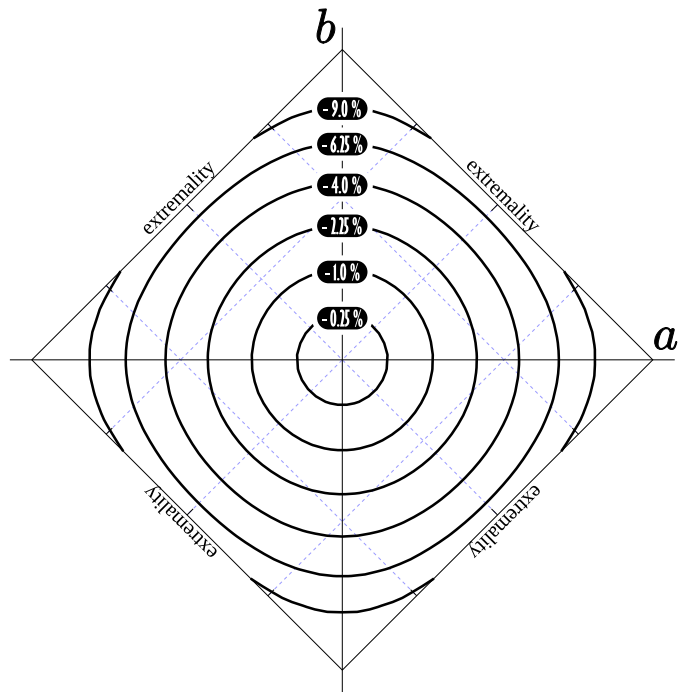, width=3.5in}}
  \caption{
    Dependence of total capture cross-section of ultra-relativistic particles
    on rotation parameters $a$ and $b$ of a five-dimensional Myers-Perry black hole.
    The capture cross-section is largest for non-rotating black hole ($a=b=0$).
    Black contour lines show relative decrease of the capture cross-section for
    rotating black holes.
  }
  \label{fig:total}
\end{figure}
\begin{figure}
  \centerline{\epsfig{file=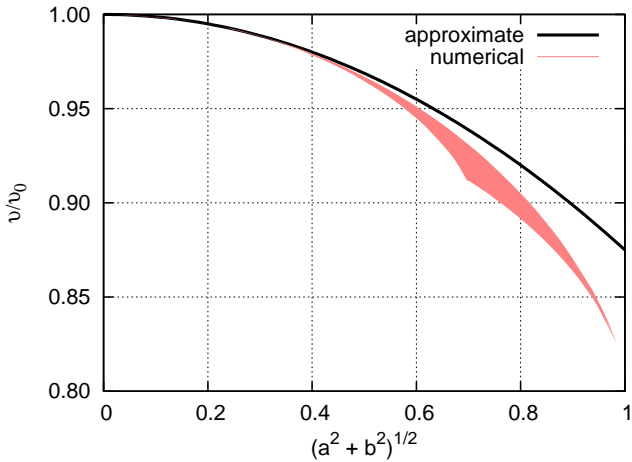, width=3.5in}}
  \caption{
    The same data for total capture cross-section of ultra-relativistic
    particles as in Figure~\ref{fig:total}, but projected on the $(a^2+b^2)^{1/2}$
    axis. As you can see, the change is second order in rotational parameters,
    and is at most $20\%$ even for near-extremal case. Thick black line shows
    results of approximation (\ref{eq:approx}), which works remarkably well.
  }
  \label{fig:approx}
\end{figure}

\begin{figure*}
  \begin{center}
  \begin{tabular}{rrr}
    \epsfig{file=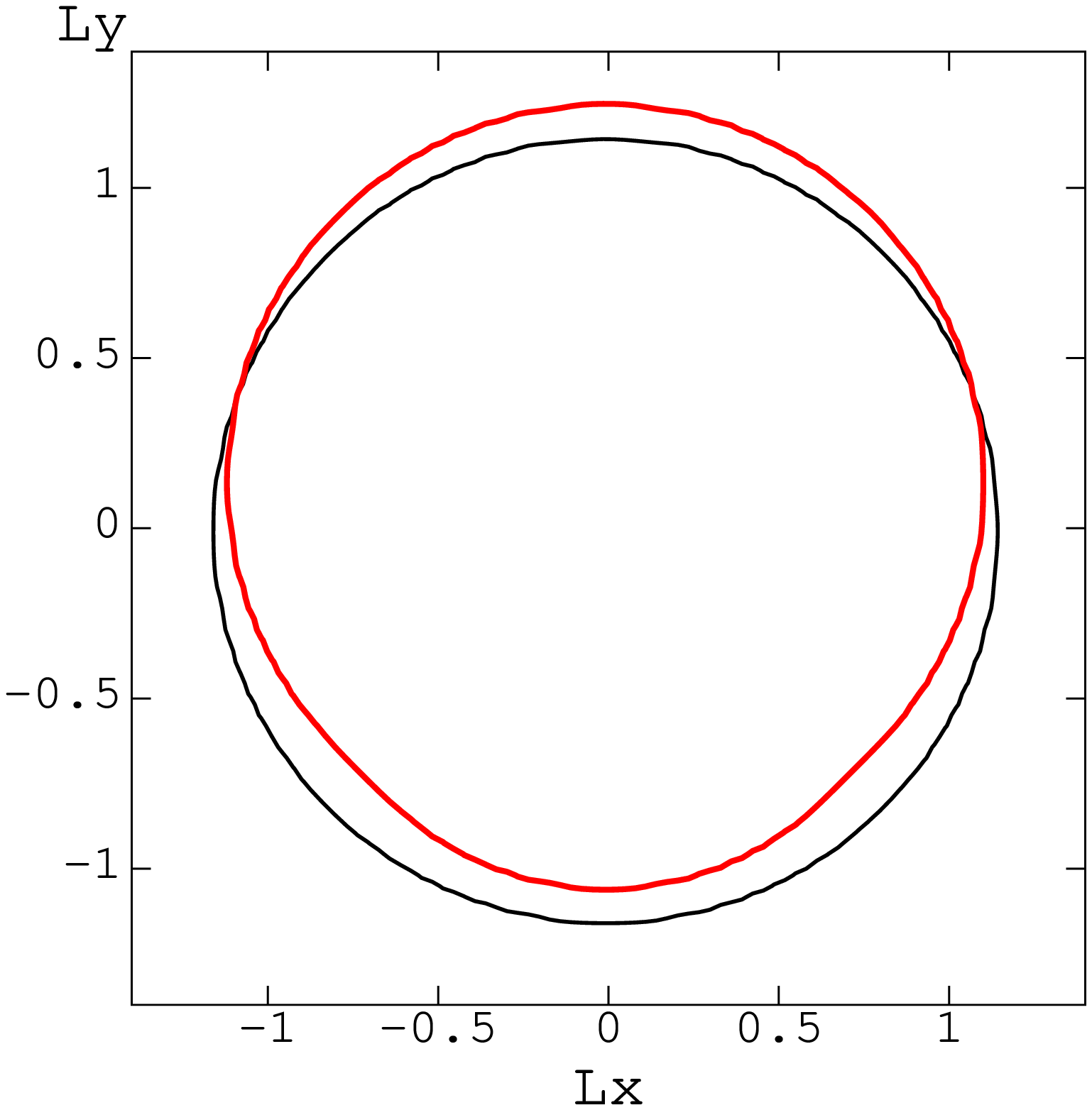, height=2.3in} &
    \epsfig{file=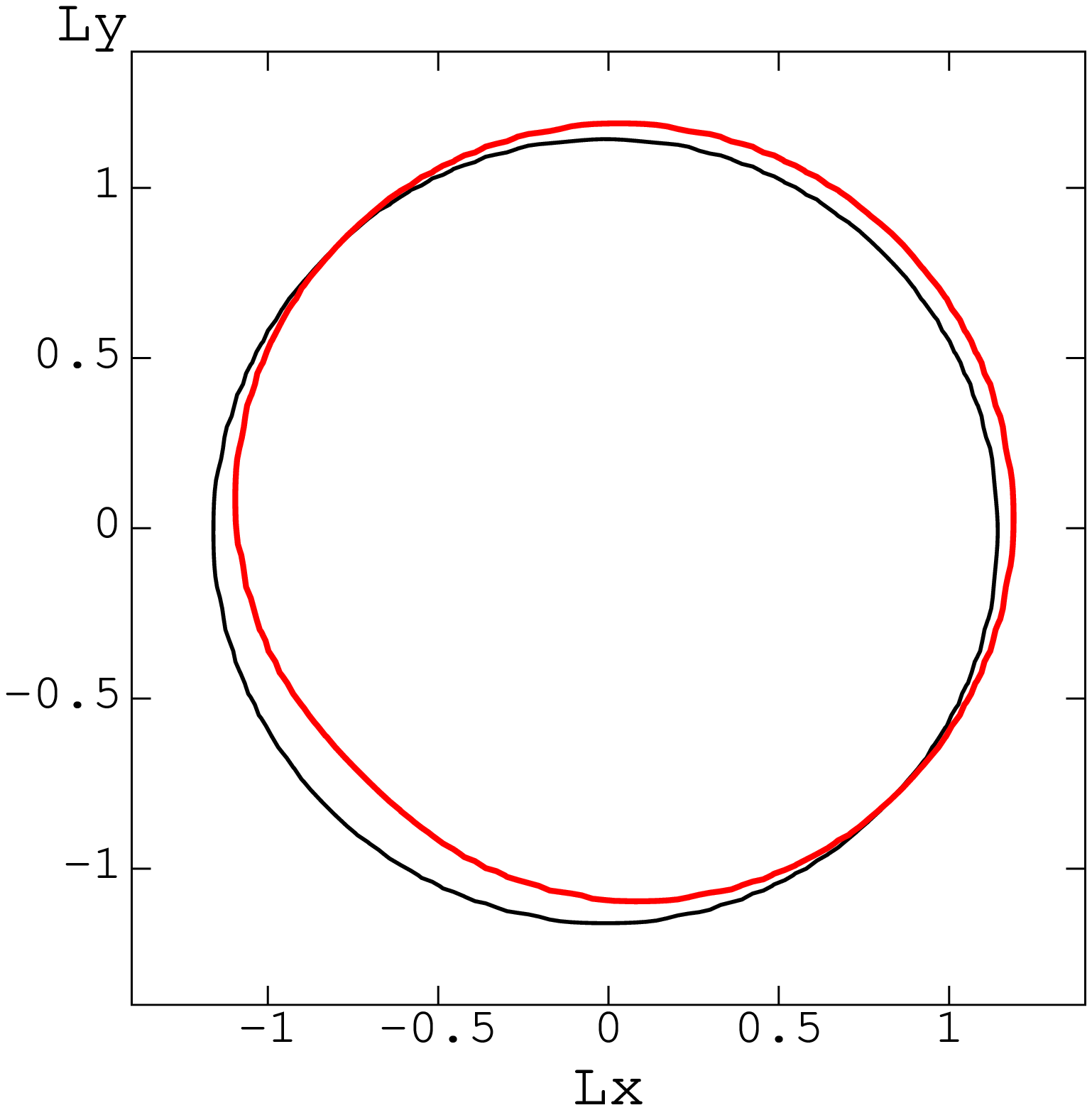, height=2.3in} &
    \epsfig{file=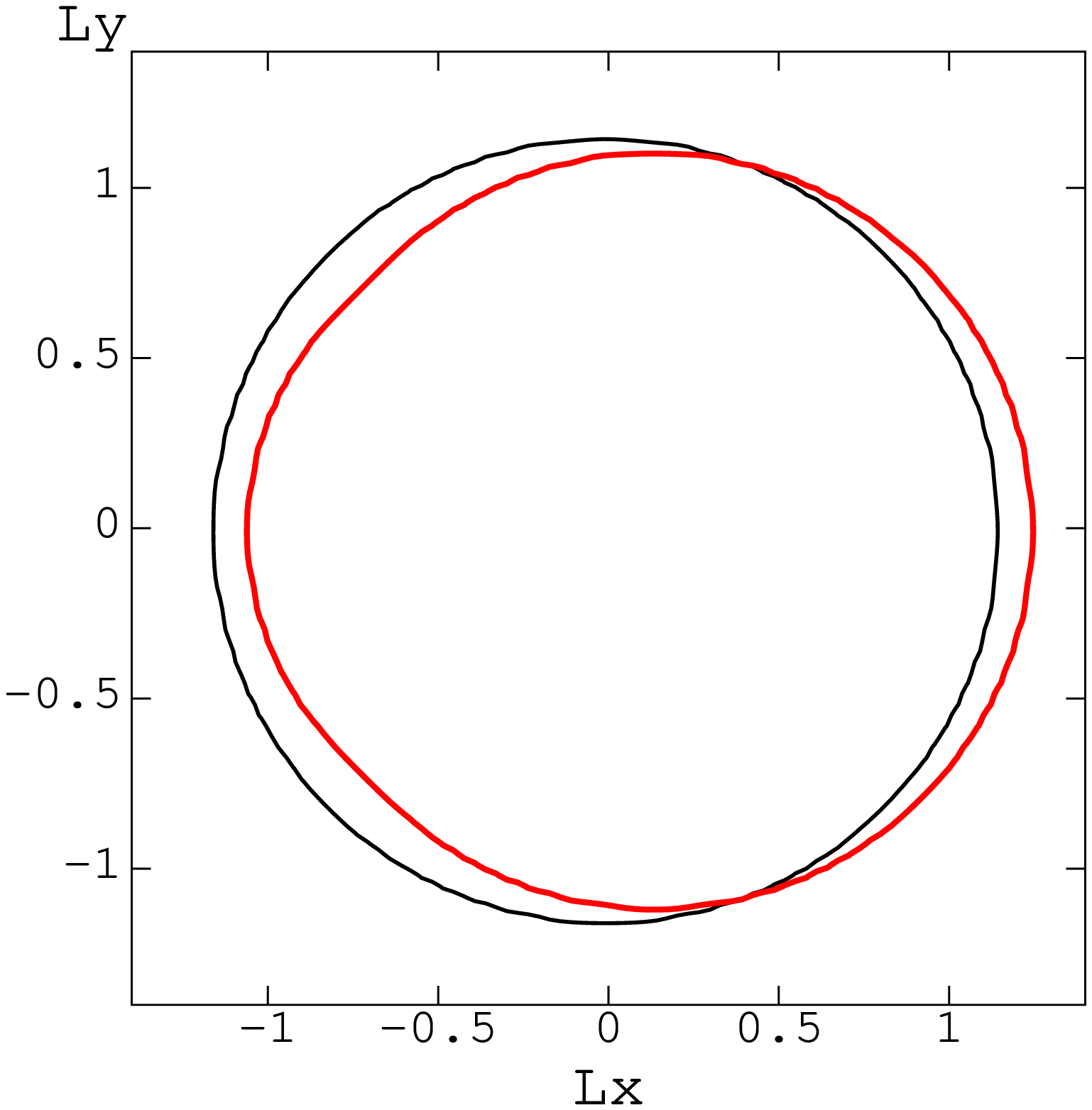, height=2.3in} \\
    \epsfig{file=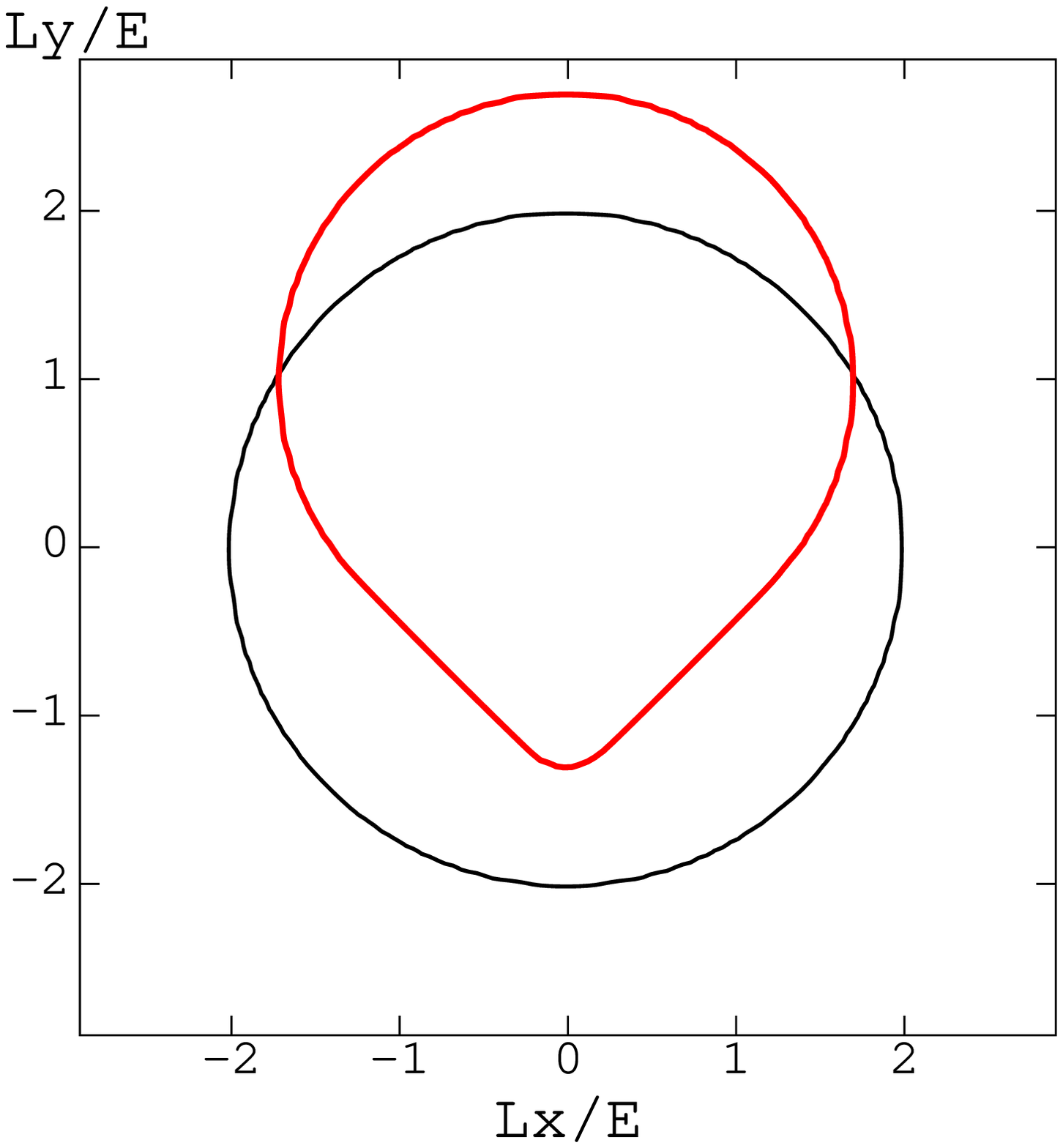, height=2.3in} &
    \epsfig{file=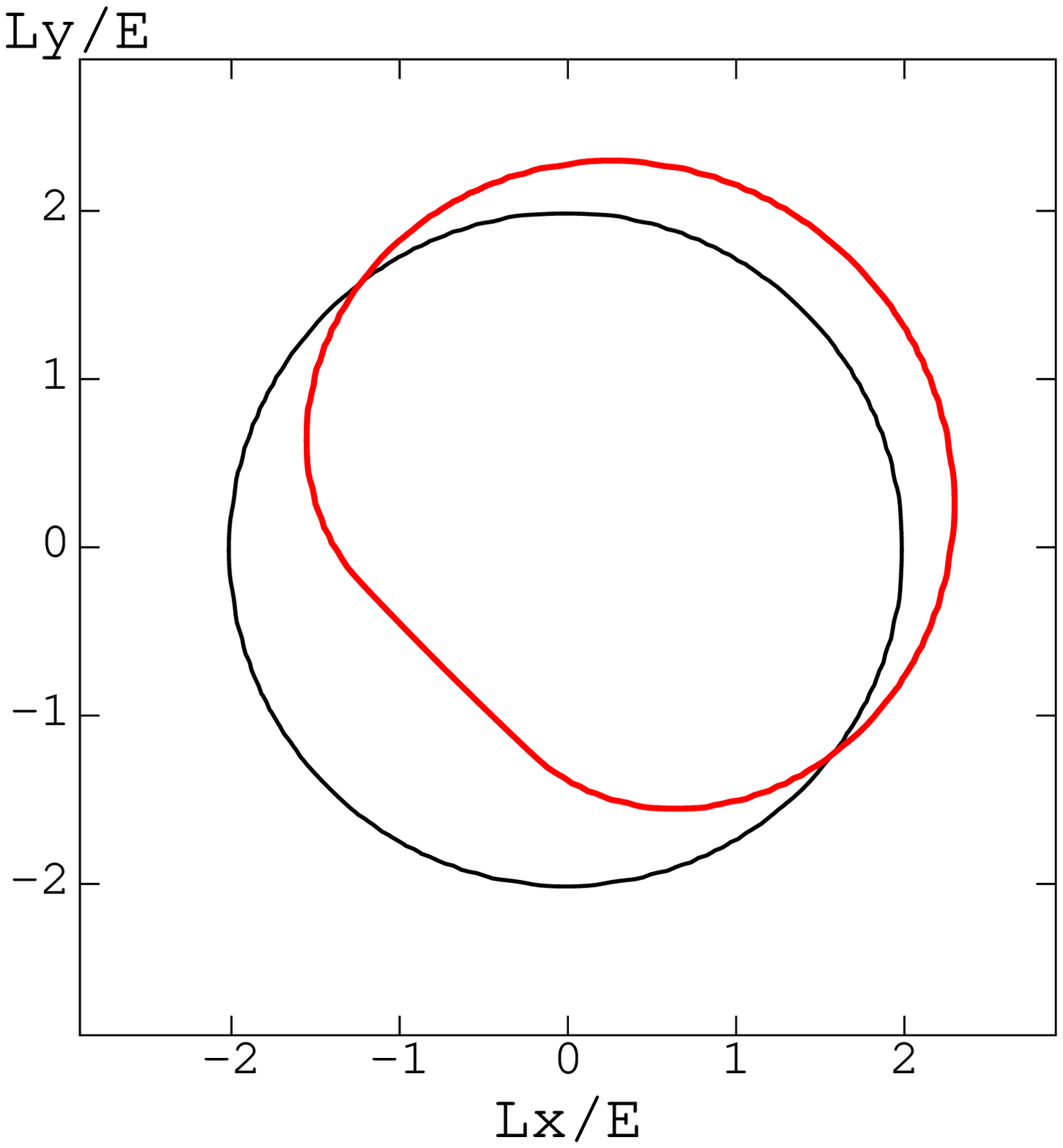, height=2.3in} &
    \epsfig{file=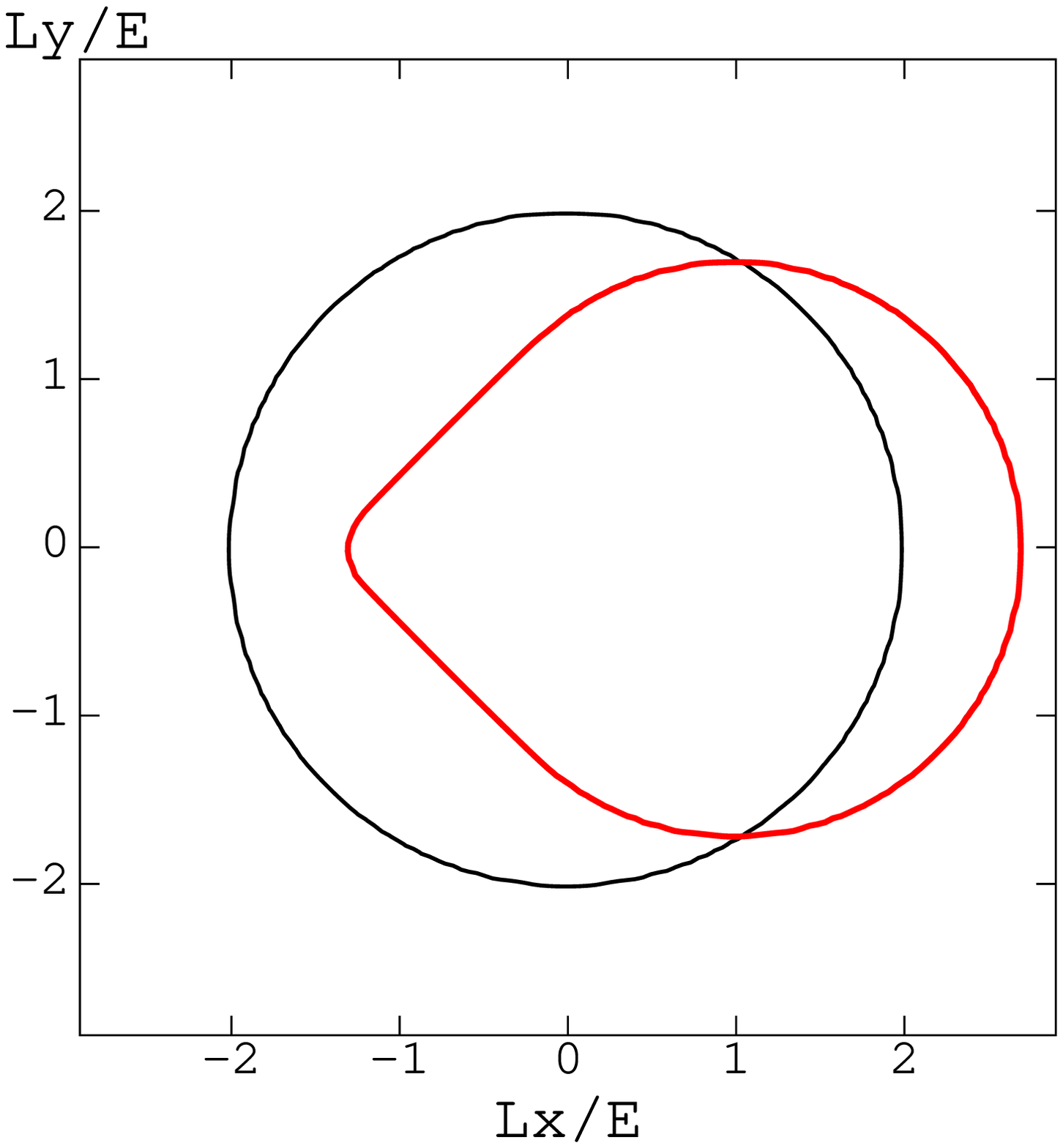, height=2.3in} \\
  \end{tabular}
  \end{center}
  \caption{
    Deformation of capture surface in near-extremal limit $(a+b)\rightarrow 1$
    with $a=0$ (left), $a=b$ (center), and $b=0$ (right) for particles with
    small initial velocities (top) and ultra-relativistic particles (bottom).
    The figure shows two-dimensional slice through the capture surface (red)
    in $L_x$--$L_y$ plane, where the deformation is most apparent. Black
    circles show the capture cross-section of a non-rotating black hole for
    comparison.
  }
  \label{fig:2D}
\end{figure*}

\begin{figure*}
  \begin{center}
  \begin{tabular}{cc}
    \epsfig{file=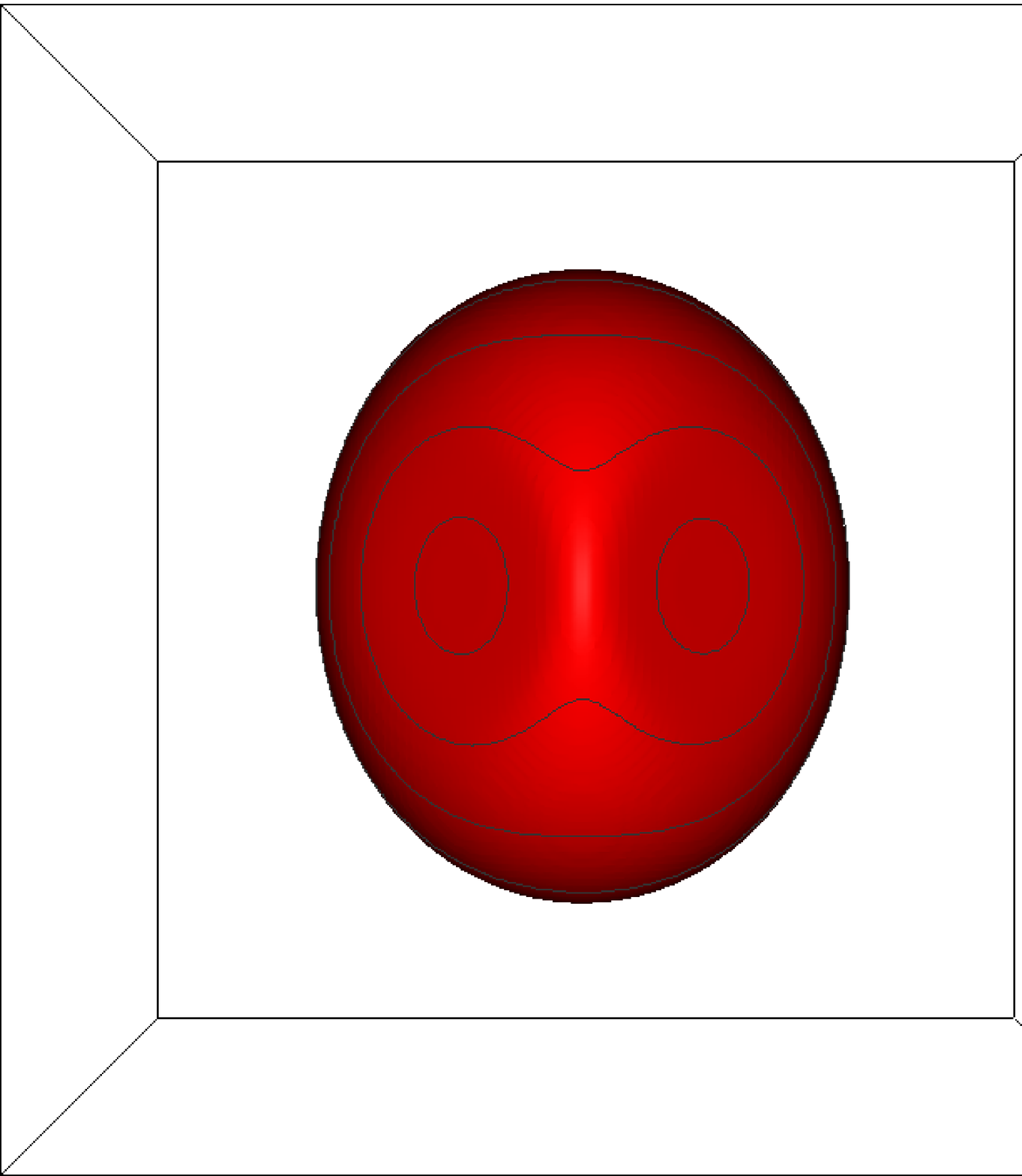, width=3.2in} &
    \epsfig{file=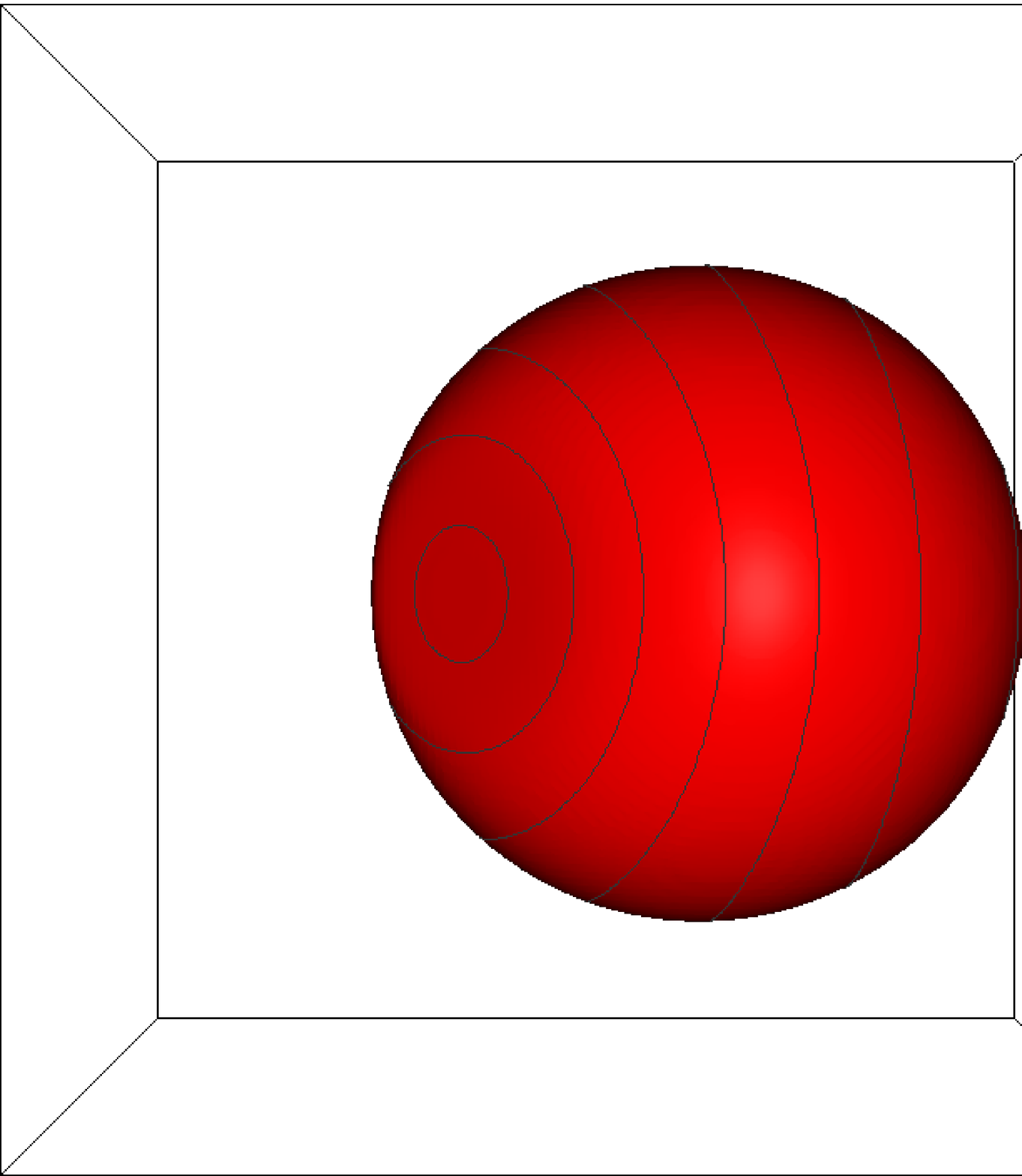, width=3.2in} \\
    \epsfig{file=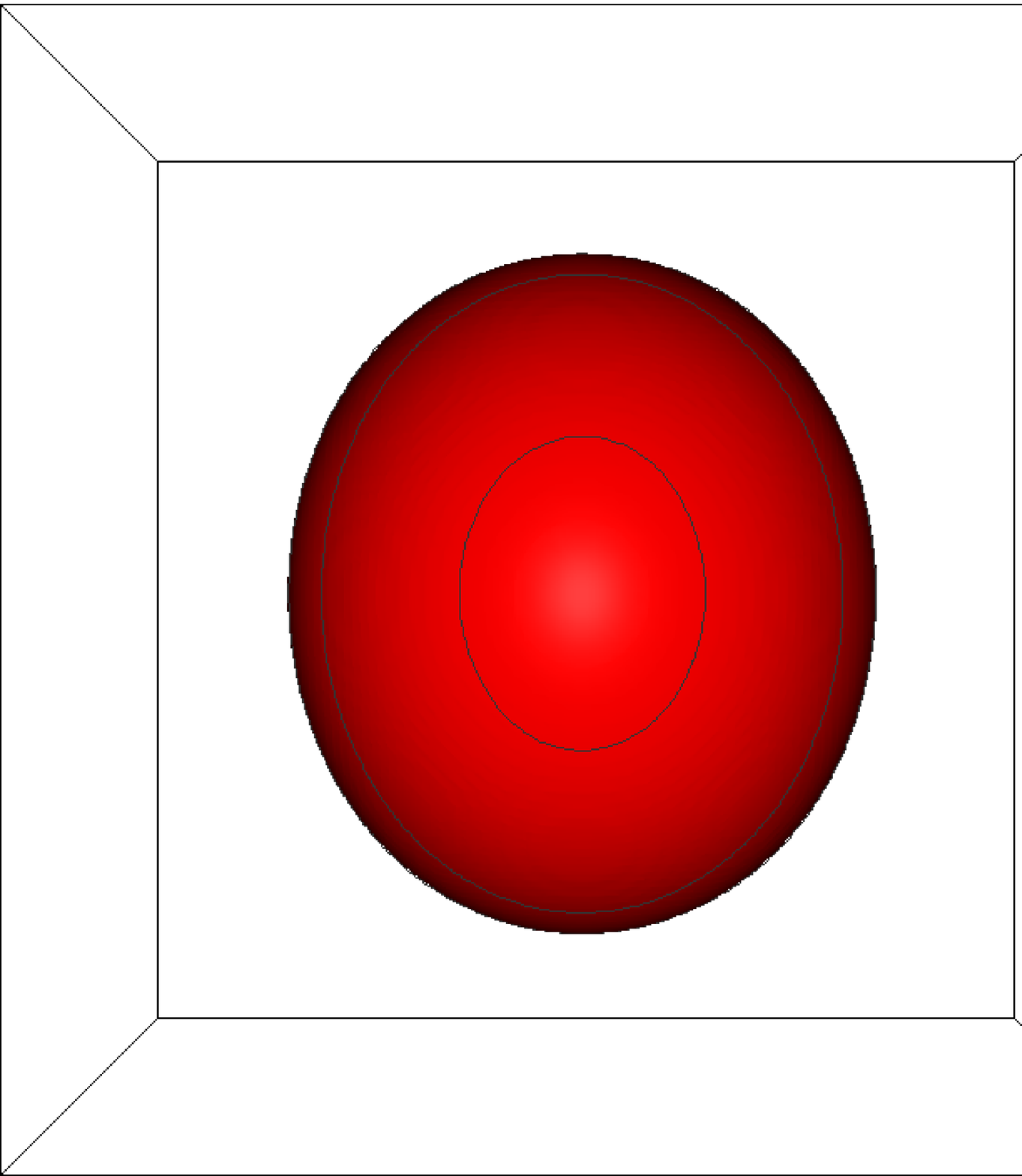, width=3.2in} &
    \epsfig{file=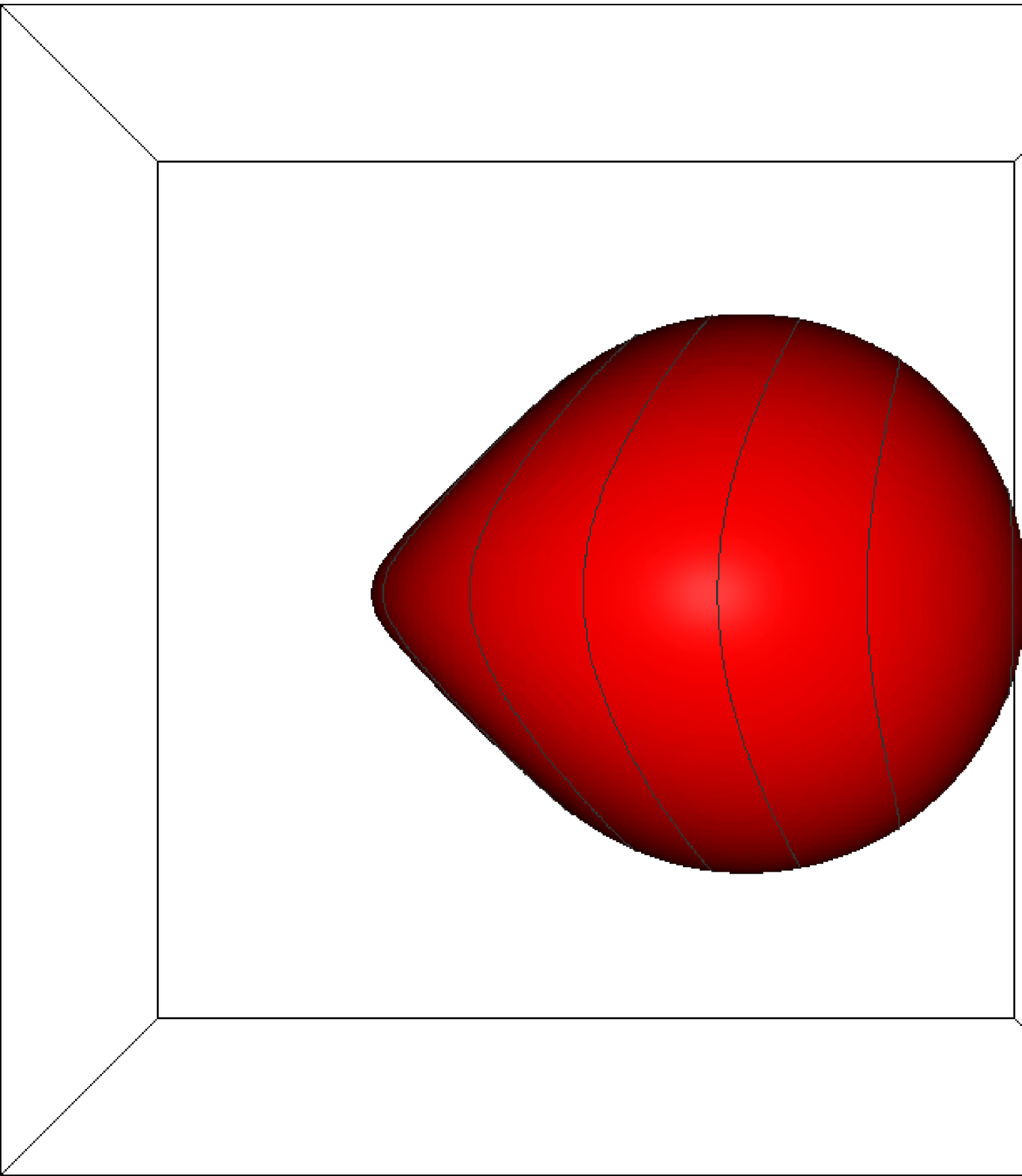, width=3.2in} \\
  \end{tabular}
  \end{center}
  \caption{
    3D rendering of the capture surface for ultra-relativistic particles
    in near-extremal limit $(a+b)\rightarrow 1$ with $b=0$
    (corresponding to 2D slice in lower right of Figure~\ref{fig:2D}).
    The four views show the capture surface (shaded red) as seen from
    $-L_x$ (top left), $-L_y$ (top right), $+L_x$ (bottom left), and
    $-L_z$ (bottom right) directions. Black contours on the surface show
    the isolines of impact parameter.
  }
  \label{fig:3D}
\end{figure*}

In the general case, solution of the algebraic equations (\ref{eq:crit})
becomes intractable, and we calculate the capture cross-sections numerically.
For the purposes of numerical evaluation, solving ordinary differential
equations is easier than dealing with a complicated system of algebraic
equations, so we determine the capture cross-section by direct integration of
the equations of motion (\ref{eq:eom}). Using a different method also allows
an independent cross-check of the analytic results presented so far.

We implement the ray-tracer using the standard fifth-order embedded
Runge-Kutta integrator with adaptive step-size control \cite{NR77}. To
determine whether the particle gets captured or not, it is only necessary to
trace the radial equation of motion, which simplifies the computations
involved. Integration starts at infinity, and stops either when the particle
reaches a turning point (i.e.~escapes), or crosses the outer horizon of the
black hole (i.e.~gets captured). The initial conditions for particle
trajectories are sampled on an uniform three-dimensional grid in impact
parameter space using scaled variable $\vec{L}/E = (p/E)\, \vec{\rho}/r_g$,
the critical values of which are finite for all particle momenta. The size of
the grid is taken to be $128^3$. To improve the quality of the capture
cross-section evaluation, adaptive mesh refinement is employed around the
capture boundary surface, with mesh refinement factor of $16$. Rays traced at
sub-grid resolution are folded back onto original grid using anti-aliasing to
produce smooth properly sampled capture surface.

We calculated the capture cross-sections as seen from different angles, for
different particle momenta, and scanned the rotational parameter space $(a,b)$
on a fine grid. Here we summarize the results of our numerical calculations,
which represent a significant computing time on a large parallel machine
(CITA's Sunnyvale cluster).

Figure~\ref{fig:total} shows the calculated dependence of total capture
cross-section of ultra-relativistic particles on rotation parameters $a$ and
$b$ of a five-dimensional Myers-Perry black hole, as seen from
$\theta_0=\pi/4$ angle. As you can see, the change in capture cross-section is
small and almost symmetric with respect to rotations in $(a,b)$-plane,
precisely as formula (\ref{eq:approx}) would suggest. This approximation is
directly compared with numerical data in Figure~\ref{fig:approx}, and the
agreement is excellent. The approximation (\ref{eq:approx}) is reliable to
$1\%$ half-way to extremality, and is not too far off even for extremal black
hole rotation.

Although an important quantity, the total capture cross-section volume does
not carry the full information about particle capture. Let us examine the
actual particle capture surfaces in more detail. Figure~\ref{fig:2D} shows a
two-dimensional slice through the capture surface in $L_x$-$L_y$ plane (where
the deformation is most apparent) for slowly moving ($E/m=1.01$) and
ultra-relativistic ($E/m \rightarrow \infty$) particles. The plots are done
for large rotational parameters $a+b \rightarrow 1$, so that extremal black
hole rotation is approached, and are as seen from $\theta_0=\pi/4$ angle. The
capture surfaces shift in the right direction (\ref{eq:shift}), but their
shapes become more complex than what could be described by a quadratic
section. From Figure~\ref{fig:2D}, it is clear that deformations of the
capture surface are most significant in the ultra-relativistic limit, which is
the reason we focused on this case for total capture cross-section.

Figure~\ref{fig:3D} shows a 3D rendering of the capture surface for
ultra-relativistic particles in near-extremal limit $a\rightarrow 1$, $b=0$.
To help visualize the surface shape, four views of the surface from different
directions are presented, with impact parameter isolines overlayed. The effect
of high rotational parameters is a quite significant deformation of the
surface, with the most prominent feature being two flattened ``cheeks''. This
flattening of the capture surface occurs for retrograde trajectories
\cite{Bardeen:1972fi}, for which the particle angular momenta $\Phi$ and
$\Psi$ have opposite signs from the black hole rotational parameters $a$ and
$b$ correspondingly. Conversely, for direct trajectories, the capture surface
``bulges'' out. This effect is not dissimilar to what happens to the the
capture circle for extremal Kerr black hole in four dimensions
\cite{1976PhRvD..14.3281Y}, except it occurs in three dimensions, and more
than one plane is involved. The flattening of the capture surface is most
likely caused by the quartic section (\ref{eq:disc}) going nearly degenerate,
but it is hard to analyze explicitly.

\section{Discussion}\label{sec:disc}

We have studied scattering and capture of particles by five-dimensional black
holes. The capture cross-section for a non-rotating black hole is bound by a perfect
two-sphere in three-dimensional impact parameter space. Rotation of the black
hole (described by two dimensionless rotational parameters $a$ and $b$)
deforms the capture surface. To linear order in rotational parameters, the
capture surface remains a sphere, but its origin shifts. This behaviour is
similar to what happens in four dimensions \cite{Barrabes:2004gn}. To second
order, the capture surface becomes an offset ellipsoid, and its volume
(i.e.~the total capture cross-section) decreases with rotation. As extremal
black hole rotation is approached, the deformation of the capture surface
becomes quite strong, and we visualize it using numerical calculations.

Although we have not considered wave propagation or quantum effects in this
work, our results on particle capture have some bearing on these more
complicated problems. Capture cross-sections of ultra-relativistic particles
have been used to estimate grey-body factors for Hawking radiation by
higher-dimensional non-rotating black holes \cite{Emparan:2000rs} using the
DeWitt approximation \cite{DeWitt:1975ys}. Although this approximation is not
entirely justified, it leads to results which agree fairly well with exact
calculations \cite{Harris:2003eg}. If the DeWitt approximation works for
rotating black holes as well, our results will provide a simple estimate for
grey-body factors for arbitrarily rotating black hole in five dimensions. The
validity of the DeWitt approximation in this case remains to be seen, however,
and it will probably not describe wave phenomena like super-radiance
adequately. Nevertheless, it might be fruitful to investigate it further.

As a final thought, we note that dependence of black hole cross-sections
\cite{Harris:2003eg}, or deflection angles \cite{Briet:2008mz}, on the
dimensionality of spacetime has been suggested as a way to determine the
number of extra dimensions (provided higher-dimensional black holes are ever
observed, of course). As the dimensionality of spacetime is increased, the
critical impact parameter for light capture decreases \cite{Emparan:2000rs},
but the dependence is not very strong. The critical impact parameter is
reduced by about $12\%$ when one goes from five to six dimensions, and less
above that. If one were to base the decision about the number of extra
dimensions on a measurement of the total capture cross-section alone, one
could be mislead by the effects of black hole rotation, which could decrease
the cross-section by comparable amount, as we show in this paper. This
degeneracy can be disentangled upon closer inspection, but serves to
illustrate the importance of black hole rotation in higher-dimensional models.

\section*{Acknowledgments}
This work was supported by the Natural Sciences and Engineering Research
Council of Canada under Discovery Grants and Undergraduate Student Research
Awards programs. Numerical computations were done on Sunnyvale cluster at
Canadian Institute for Theoretical Astrophysics.



\end{document}